\documentclass[showpacs,prc,preprint,groupedaddress,amsmath,amssymb]{revtex4}
\usepackage{graphicx}% Include figure files

\begin{document}

%\preprint{}

\title{The dissociation of the $J/\psi$ by light mesons and chiral symmetry}

\author{Alex Bourque}
\author{Charles Gale}

%\thanks{}
\affiliation{Department of Physics, McGill University \\ 3600 University Street, Montreal, QC,
Canada H3A 2T8}


\date{\today}

\begin{abstract}
The implication of chiral symmetry for the pion-induced dissociation 
of the $J/\psi$  is examined in detail. It is shown how the low-energy dynamics of pions,  constrained by chiral symmetry, affect the dissociation cross--section. The derived soft--pion theorem is then integrated into a
Lagrangian model which includes also abnormal parity content and chiral--symmetric form factors.
Dissociation by the $\rho$ meson is also considered.
\end{abstract}

\pacs{13.75.Lb, 11.30.Rd, 12.38.Mh}
%\keywords{}

\maketitle

\section{Introduction}
It is predicted that at very high energy densities, confined hadronic matter melts into a novel form: 
the quark--gluon plasma (QGP). Several signatures to characterize its properties 
within the context of heavy ion collisions have been
proposed. One of these, initially championed by Matsui and Satz
\cite{Mat86}, is charmonium suppression. 
It rests on the observation that correlated $\bar c c$ pairs 
created in the earliest stage of the collisions through hard
scatterings probe all subsequent stages of the system evolution. In
particular, if a QGP is formed, they argued that 
the observed yield should be suppressed because of color screening \cite{Mat86}. The current view not only includes suppression, but also the 
regeneration of charmonium  \cite{The06,Gra04,And03}. Moreover, recent lattice data suggests that the $J/\psi$ may survive in the plasma well above $T_C$ \cite{Asa04,Dat04}, implying that
that there could be no direct QGP suppression of this meson \cite{Kar06}. See however Ref. \cite{Mocsy}. 

Before a claim of any definite QGP effects is made, it is essential to verify that
the results cannot be reproduced by more mundane nuclear effects. Of all possible
mechanisms, 
charmonium dissociation by nucleons is probably the most important one. 
Indeed, it is seen to be sufficient to explain the suppression patterns observed at the SPS not only 
for $p+A$ systems, but also $O+U$ and $S+U$  collisions \cite{Ger92}. For heavier systems, 
nuclear suppression is not sufficient to account for experimental
observations. For example, in $Pb+Pb$ collisions at SPS, 
an abnormal suppression is observed. One possible cause of the charmonium suppression
could of course be screening \cite{Kha97}. But  dissociation by light--meson co--movers can 
also go a long way in explaining the observed data \cite{Fta88,Gav88,Vog88,Bla89_2,Arm98,Arm99,Cap03,Cas97,Cas97_2,Spi99,Bar03}. 

In most phenomenological studies, the dissociation cross--section by co--movers is a model parameter,
and little is said about the underlying microscopic mechanisms. Since experimental information
about dissociation processes is scarce, one has to rely on theoretical studies.
Several approaches are possible. One model calculates the dissociation cross--sections by using 
constituent quarks and a non--relativistic potential \cite{Bar03,Mar95,Won00,Won02}. The
dissociation processes then arise through the exchange of quarks. 
A fully relativistic constituent quark model
 can also be constructed based on
an extension of the Nambu--Jona Lasino (NJL) model to the charm sector
\cite{Dea98,Pol00,Dea03,Mai04,Mai05,Lap06}. 
Dissociation then occurs through quark-- and meson--exchanges. 
Being non--renormalizable, the model requires the specification
of an ultraviolet loop cutoff. One can circumvent the need of such a
cutoff by introducing form factors at the quark level. This leads to the extended non--local NJL
model of Ref.~\cite{Iva05}. Another model relies rather on extrapolations of QCD sum rule (QCDSR) results
to extract momentum dependent vertices \cite{Dur03,Nav00,Bra01,Nav02,Nav02_2,Mat02,Dur03_2,Aze04}.
Finally, phenomenological Lagrangians \cite{Mat98,Lin00,Lin00_2,Hag00,Hag01,Oh01} can be utilized. There, in order to account for the composite nature of the mesons,
ad--hoc form factors are often introduced. 

These models then produce cross--sections ranging from sub--millibarn to a few
millibarns. Moreover, their energy behaviour can be quite different \cite{Rap04}.
This is compounded by the fact that in many models, chiral symmetry is not clearly implemented. As pointed
out in Ref.~\cite{Nav01} in the context of the Lagrangian--type models, chiral symmetry implies that for
the normal parity content of the process $J/\psi + \pi \rightarrow (D^* + \bar D)+ (\bar D^* + D)$ the pion should decouple 
in the soft--momentum limit leading to a vanishing amplitude. Since this process is
considered to be dominant, owing to the abundance of pions, quantifying this effect is therefore
important.

In Ref.~\cite{Bou05_1}, the effect of implementing chiral symmetry in a simple Lagrangian model
without form factors was considered in contrast with previous phenomenological Lagrangians \cite{Mat98,Lin00,Lin00_2,Hag00,Hag01,Oh01}. It was shown there that for the
$J/\psi + \pi \rightarrow (D^* + \bar D)+ (\bar D^* + D)$ process a reduction at threshold did occur. 
It is the purpose of this article to propose an improved Lagrangian model that
incorporates not only chiral symmetry and form factors, but also other dissociation channels, i.e.,
the so--called abnormal parity processes \cite{Oh01}. The $\rho$--induced 
dissociation cross--sections will also be evaluated in order to assess the relative importance of
dissociation by other light resonances. 

This article is organized as follows: we first discuss the soft--pion theorem. The relevant
degrees of freedom are then introduced, and these enable us to write down   
chiral Lagrangian densities. Inelastic
cross--sections are extracted and the soft--pion theorem is explicitly verified. Once parameters are fixed and
symmetry preserving form factors are introduced, the relative strengths of
chiral symmetry, abnormal parity content, and $\rho$--dissociation effects on the cross--sections are
presented and discussed.

\section{\label{decoupling}Decoupling of pions in the soft--momentum limit}
First consider the case where the chiral symmetry is exact. The axial current for
the Goldstone realization of the chiral symmetry \cite{Mos99} is
\begin{equation}
{A}_\mu(x) = f^0_\pi\partial_\mu{\pi} + \dots
\end{equation}
where $f^0_\pi$ is the pion decay constant in the chiral limit. 
Using the LSZ reduction  formulae \cite{Pes95} and following Weinberg
\cite{Wei95}, the expectation value of the 
current between an arbitrary in-- and out--state becomes 
\begin{equation}
\int d^4x \left<\alpha\right|{A}_\mu(x)\left|\beta\right> e^{-ip\cdot x}
= \frac{p^\mu f^0_\pi}{p^2}i\mathcal{M}^{0\pi}_{\beta \rightarrow \alpha} +
\mathcal{N}^\mu_{\beta\rightarrow\alpha}
\end{equation}
where $\mathcal{M}^{0\pi}_{\beta \rightarrow \alpha}$ is the transition
amplitude in the chiral limit for the absorption of an incoming pion with momentum 
$p$,  and $\mathcal{N}^\mu_{\beta\rightarrow\alpha}$
are the regular terms near the pion pole. Contracting the pion momentum on both
side yields the current conservation condition:
\begin{equation}
\left<\alpha\right|p_\mu{A}^\mu(p)\left|\beta\right> = 
f^0_\pi i\mathcal{M}^{0\pi}_{\beta \rightarrow \alpha}  + 
p_\mu\mathcal{N}^\mu_{\beta\rightarrow\alpha} = 0
\end{equation}
Under the assumption that $\mathcal{N}^\mu_{\beta\rightarrow\alpha}$ is regular
near the pion pole, the pion then decouples in the soft--momentum limit giving
\begin{equation}
\mathcal{M}^{0\pi}_{\beta \rightarrow \alpha} \rightarrow 0.
\end{equation}
This constraint, first studied by Adler \cite{Adl65}, is an example of 
how the chiral symmetry manifests itself in the Goldstone mode for low--energy 
scattering. A general proof with many pions can be obtained \cite{Wei95}.

Knowing that chiral symmetry is only partially conserved, let us now consider
how the above theorem  is modified. Under the PCAC hypothesis \cite{Mos99},
the current matrix elements now become
\begin{equation}
\left<\alpha\right|{A}^\mu(p)\left|\beta\right> = 
\frac{p^\mu f_\pi}{p^2-m_\pi^2}i\mathcal{M}^\pi_{\beta \rightarrow \alpha} +
\mathcal{N}^\mu_{\beta\rightarrow\alpha}
\end{equation}
where $f_\pi$ is the decay constant for an explicitly broken chiral symmetry.
Assuming that the explicit chiral breaking occurs only through 
a $m_\pi^2$ dependence the pion--absorption transition amplitude reduces to
\begin{equation}
\lim_{p \rightarrow 0} \mathcal{M}^\pi_{\beta \rightarrow \alpha} =
\lim_{p \rightarrow 0} \mathcal{M}^{0\pi}_{\beta \rightarrow \alpha} \rightarrow 0.
\end{equation}
This is a strong version of the smoothness 
assumption \cite{Don94,Sak69} which requires that the amplitude does not
change significantly from $p^2 = m_\pi^2$ to $p^2 =0$. 

\begin{figure}[!htp]
\begin{center}
\includegraphics[scale=0.7]{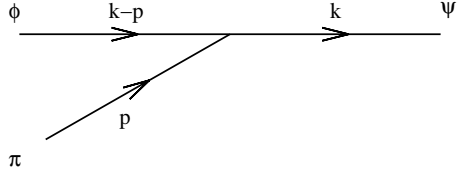}
\caption{Exception to the decoupling theorem due to a kinematic singularity.}
\label{exception}
\end{center}
\end{figure}

The above two derivations assume that no other singularities exist besides that
provided by the pion pole, or in other words, that $\mathcal{N}^\mu_{\beta\rightarrow\alpha}$ is
regular. This is in general not true \cite{Wei95,Sak69}. Fig.~\ref{exception} shows the basic 
sub--diagram where an initial off--shell particle absorbs a pion and then emits an 
on--shell particle. These 
two particles could for example be mesons. If the two particles 
have the same mass, then a kinematical singularity will develop in the 
soft--pion limit, i.e., the denominator of the incoming particle propagator becomes
 \begin{displaymath}
\lim_{p \rightarrow 0 } (k-p)^2-M^2 = \lim_{p \rightarrow 0 } k^2 -2k\cdot p
+p^2 -M^2 \rightarrow  k^2 - M^2 \rightarrow 0
 \end{displaymath}
 where $M$ is the mass of the two particles and $k^2 = M^2$ since the outgoing
 particle is on--shell. 
 
This singularity can occur in two cases. First, when the incoming and outgoing 
particles have degenerate masses as it is sometimes realized for a
chirally restored vacuum (e.g., $\sigma$-- and $\pi$--mesons are degenerate).
 The other case manifests itself for abnormal parity 
interactions which permit the absorption/emission of a pion from a single particle.
An example of such a process is between a pion and an 
isospin--doublet vector meson with negative parity, $V$, i.e., 
\begin{equation}
\mathcal{L}_{\pi VV} = g\epsilon^{\mu\nu\alpha\beta}  \partial_\mu V_\nu
\tau\cdot{\pi}
\partial_\alpha V^\dagger_\beta
\end{equation} 
where $\epsilon^{\mu\nu\alpha\beta} $ is the four--dimensional anti--symmetric
tensor. This interaction will then generate a singularity for soft--pion
kinematics since the incoming and outgoing particles are identical.

\section{Degrees of freedom and chiral symmetry}
We now wish to build a chirally invariant Lagrangian. Doing so will ensure that the soft--pion limit
is exhibited by the model. The principal difficulty is to identify the relevant degrees of
freedom. In the final stage of a heavy ion collision, the relative momentum of the
$J/\psi$ and a light meson is of the order of a few GeV: the charmonium dissociation is thus expected to be
dominated by those processes with the smallest excitation threshold, i.e.,
cross--sections with the lowest--mass final states. Therefore, it is sufficient
to consider the dissociation processes into the lowest--mass open charmed
mesons resulting from the interactions between the $J/\psi$, $D$, $D^*$
and the light mesons.
 
This point is incorrect if chiral symmetry is to be maintained. Indeed,
as pointed out in Ref.~\cite{Bou05_1} inclusion of the chiral partners is essential
for the decoupling theorem to hold. It is thus expected that the 
chiral partners, even though they do not  appear in the final states, 
still can play an important role through exchange diagrams. 
With this in mind, identifying the chiral partners of the $D$ and $D^*$ mesons 
is essential. Since they are pseudo--scalar and vector mesons, respectively, 
their chiral partners are expected to be a scalar and an axial--vector
particles. Moreover, under the heavy--quark spin symmetry, they should have 
similar masses. We see from Ref.~\cite{Yao06} that
the $D^*_0$ and $D_1$ mesons are candidates for the scalar and axial partners,
respectively \footnote{Ref. \cite{Yao06} states that the quantum numbers of the $D^*_0$ and $D_1$ mesons have to be confirmed.}.

Introducing the chiral partners amounts to having a linear realization of chiral symmetry.
One could also decide not to introduce these 
additional mesons, and consider a non--linear realization 
of chiral symmetry by letting, for example, the chiral partner masses go 
to infinity. The $D$ and $D^*$ would have non--linear 
transformation properties under the axial sub--group. This
approach is the one used in building the Lagrangians incorporating heavy--quark
spin--flavor symmetry \cite{Pol00}. For this study, the linear representation will be used.  The
open charmed mesons  
will be then the  $D$, $D^*$, $D^*_0$, and $D_1$. 

In order to build a chiral invariant Lagrangian, it is convenient to define chiral fields. In the Appendix \ref{field_rep}
these are identified by considering the various possible quark bi--linears. Knowing the transformation properties
under chiral symmetry of the light and heavy quark then permit to determine that of the mesons. The
chiral fields are then found to be
\begin{eqnarray}
W &=& \sigma + i \pi,\\
W^\dagger &=& \sigma - i \pi,\\
A_{R,L} &=& \rho \pm a_1, \\
D_{R,L} &=& D^*_0 \pm iD, \\
D^*_{R,L} &=& D^* \pm D_1 
\end{eqnarray}
where $W$ and $A_{R,L}$ are isospin triplets, and $D_{R,L}$ and $D^*_{R,L}$ are isospin doublets.

\section{\label{densities}Lagrangian densities}
We first write down the free field Lagrangian by defining the following field strengths:
\begin{eqnarray}
F_{R,L}^{\mu\nu} &=& \partial^\mu A^\nu_{R,L} - \partial^\nu A^\mu_{R,L}
\end{eqnarray} 
for an arbitrary left-- and right--handed vector field. Then, starting from the linear sigma model, 
the free field Lagrangian reads
\begin{eqnarray}
\mathcal{L}_0 &=& \frac{1}{4} Tr\left[\partial_\mu W \partial^\mu W^\dagger \right] -\frac{1}{4}\mu^2
Tr\left[WW^\dagger\right]  + \frac{f_\pi m_\pi^2}{4} Tr\left[W + W^\dagger\right]  - \frac{1}{16}Tr\left[F^L_{\mu\nu}F^{\mu\nu}_L +  F^R_{\mu\nu}F^{\mu\nu}_R \right] \nonumber \\
&+&\frac{m_0^2}{4}Tr\left[A_{L\mu}A^{\mu}_L+A_{R\mu}A^{\mu}_R\right] + \frac{1}{2}\left(\partial_\mu D_L \partial^\mu \bar D_L + \partial_\mu D_R \partial^\mu \bar D_R \right) -  \frac{M^2}{2}\left(  D_L\bar D_L + D_R \bar D_R\right) \nonumber \\
&-& \frac{1}{8}\left(F_{\mu\nu}^{D_L^*}F^{\mu\nu}_{\bar D_L^*}+F_{\mu\nu}^{D_R^*}F^{\mu\nu}_{\bar D_R^*}
\right)  + \frac{M^{*2}}{2}\left(D^{*\mu}_L\bar D^*_{L\mu} + D^{*\mu}_R \bar
D^*_{R\mu}\right)
\end{eqnarray}
where $M$ and $M^*$ are the degenerate masses of open charmed mesons and $m_0$ that of
the $\rho$ and $a_1$ mesons. Degeneracies will be lifted by spontaneous 
chiral symmetry breaking once the interactions are included, as in the linear sigma model
\cite{Mos99},
 which will result in mass splittings between  $D$ and $D_0^*$, and $D^*$
and $D_1$. A pion mass has also been included with the
third term, and thus chiral symmetry is explicitly broken.

For the interactions, the working assumption here will be that only the three-- and four--point 
interactions with the lowest number of derivatives are to be considered 
\footnote{This assumption is strictly valid only if all terms 
with higher powers of derivative are suppressed for the considered kinematical 
regime \cite{Don94}.}.
Since the Lagrangian density is of dimension four and the mesonic fields are of
dimension one, the three--point interactions will have couplings scaling as 
$M^{1-n}$ where $M$ is an arbitrary mass--scale and $n$ is the number of 
derivatives, while the four--point interactions having one more field operator 
will scale as $M^{-n}$. Furthermore, only the minimal interactions with the chiral partners of the $D$ 
and $D^*$ mesons will be added to maintain chiral symmetry. Practically, this
implies that all the interactions with $D^*_0$ and $D_1$ fields will be
generated by the spontaneous chiral symmetry breaking. Moreover, only the three-- and four--point 
interaction terms necessary to contruct the amplitudes with the considered final states are explicitly written down. Finally, aside from the requirement that the Lagrangian density be real,
the other tools used to construct the phenomenological 
Lagrangian are parity and charge conjugation invariances (which are valid
symmetries of QCD). The effects of these 
discrete transformations on the field content are listed in Appendix \ref{interactions} 
as well as the resulting interactions.

The next step is to make explicit the spontaneous chiral symmetry breaking by
shifting the $\sigma$ field in $W$ by $\sigma \rightarrow \sigma + \sigma_0$
as in the linear sigma model. Doing so yields 
the new free field Lagrangian
\begin{eqnarray}
\mathcal{L}_0 &=& \frac{1}{2}\partial_\mu \pi \partial^\mu \pi - m_\pi^2
\pi^2
+\frac{1}{2}\partial_\mu\sigma\partial^\mu\sigma-m_\sigma^2\sigma^{2}
 - \frac{1}{8} Tr\left[ F^{\mu\nu}_\rho F_{\mu\nu}^\rho\right]
+\frac{1}{4} m_0^2 Tr \left[\rho_\mu ^2\right] \nonumber \\
&-& \frac{1}{8} Tr\left[ F^{\mu\nu}_{a_1} F_{\mu\nu}^{a_1}\right]  +\frac{1}{4} m_0^2
 Tr \left[a_{1\mu} ^2\right] 
+\partial_\mu D \partial^\mu \bar D -\left(M^2-2\Delta
\sigma_0\right)D\bar D + \nonumber \\
&+& \partial_\mu D^*_0 \partial^\mu \bar D^*_0 - \left(M^2 + 2\Delta
\sigma_0\right) D^*_0 \bar D^*_0 
- \frac{1}{4}F^{D^*}_{\mu\nu}F^{\mu\nu}_{\bar D^*} + \left(M^{*2} -
2\Delta^*\sigma_0\right)D^*_\mu\bar D^{*\mu} \nonumber \\
&-&\frac{1}{4}F^{D_1}_{\mu\nu}F^{\mu\nu}_{\bar D_1} + \left(M^{*2} +
2\Delta^*\sigma_0\right)D_{1\mu}\bar D^{\mu}_1 
+ ig^{(0)}_{WDD^*}\sigma_0 \left(\partial_\mu D^*_0 \bar D^{*\mu} -
D^{*\mu}\partial_\mu \bar D^*_0 \right)  \nonumber \\
 &+& g^{(0)}_{WDD^*}\sigma_0 \left(\partial_\mu D \bar D_1^{\mu} -
D^{\mu}_1\partial_\mu \bar D \right)
\end{eqnarray} 
where the expressions for $m_\pi$ and $m_\sigma$ are the same as for the linear sigma model \cite{Mos99}, and
$g^{(0)}_{WDD^*}$, $\Delta$, and $\Delta^*$ are coupling constants.
We note that the introduction of interactions generate mass splittings between 
the $D$ and $D_0^*$ mesons, and between the 
$D^*$ and $D_1$ mesons respectively; thus lifting the mass degeneracies. The
$D$ meson masses then read
\begin{eqnarray}
m_D^2 &=& M^2 - 2\Delta\sigma_0,\quad m_{D^*_0}^2 = M^2 + 2\Delta\sigma_0, \nonumber \\
m_{D^*}^{2} &=& M^{*2} - 2\Delta^*\sigma_0,\quad m_{D_1}^2 = M^{*2} + 2\Delta^*\sigma_0. \nonumber 
\end{eqnarray}
Moreover, the introduction of the interactions induces mixing between $D^*_0$ and $D^*$ fields, and 
between $D$ and $D_1$ fields.  
To cast the Lagrangian into a canonical form would thus require making
field redefinitions. These are involved and would lead to additional
interactions with higher powers of momentum, which is contrary to the original
assumption of limiting possible interactions to those with the lowest powers of
momentum. Moreover, the non--chiral invariant model of 
Ref.~\cite{Oh01} with which we wish to make comparison has no such mixings. For this
study, the coupling constant $g^{(0)}_{WDD^*}$ is thus set to zero removing the
mixing. Finally, in this model, the $\rho$ and $a_1$ mesons have degenerate masses. This is of
no importance here since we wish to compute only the cross--sections with the
two lightest mesons, namely the $\pi$ and $\rho$. In Appendix \ref{interactions}, the relevant interactions for the $J/\psi$ meson by a pion or a $\rho$ meson  
are listed. They include  normal and abnormal interactions. As discussed in Section
\ref{decoupling}, the latter are expected to circumvent the low--energy theorem.

\section{Inelastic scattering amplitudes}
All amplitudes discussed in this section are explicitly written down in Appendix \ref{amplitudes}. 
\subsection{$\pi +J/\psi$}
\begin{figure}[!htb]
\begin{center}
%\scalebox{0.55}[0.55]{\includegraphics{PIJ.eps}}
\includegraphics[scale=0.60]{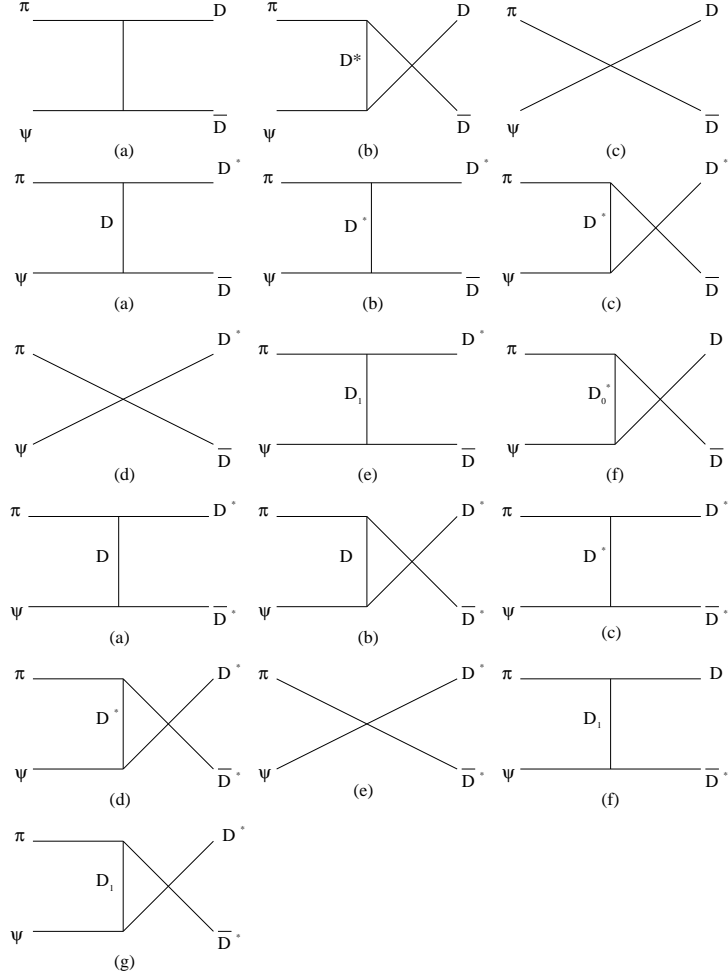}
\caption{Diagrams for $\pi +\psi$ dissociation}
\label{MPIJ}
\end{center}
\end{figure}

The pion--dissociation of the $J/\psi$  proceeds through three processes, namely:
\begin{eqnarray}
\mathcal{M}_{1} &=& \sum_{i} \mathcal{M}^\rho_{1i} \epsilon_\rho \left(p_\psi\right), \\
\mathcal{M}_{2} &=& \epsilon^*_\mu \left(p_{D^*}\right) \sum_{i} \mathcal{M}^{\mu\rho}_{2i} \epsilon_\rho \left(p_\psi\right) , \\
\mathcal{M}_{3} &=& \epsilon^*_\mu \left(p_{D^*}\right)\epsilon^*_\nu
\left(p_{\bar D^*}\right)\sum_{i} \mathcal{M}^{\mu\nu\rho}_{3i}\epsilon_\rho
\left(p_\psi\right)
\end{eqnarray}
where $\epsilon_\rho \left(p_\psi\right)$, $\epsilon^*_\mu
\left(p_{D^*}\right)$, and $\epsilon^*_\nu
\left(p_{\bar D^*}\right)$ are the polarization vectors for the $J/\psi$, $D^*$
and $\bar D^*$ mesons respectively. The first and last amplitudes arise only due to abnormal parity
interactions, while $\mathcal{M}_{2}$ contains one abnormal parity exchange process ((b) in Fig.~\ref{MPIJ}).
Note also that the amplitude for the final state $\bar{D}^*D$ is
obtained from the conjugate of amplitude $\mathcal{M}_{2}$.
\subsection{$\rho +J/\psi$}
For the $\rho$--meson induced dissociation, three processes are examined:
\begin{eqnarray}
\mathcal{M}_{4} &=& \sum_{i} \mathcal{M}^{\delta\rho}_{4i} \epsilon_\rho
\left(p_\psi\right) \epsilon_\delta \left(p_{\rho}\right)  \\
\mathcal{M}_{5} &=& \epsilon^*_\mu \left(p_{D^*}\right) \sum_{i}
\mathcal{M}^{\mu\delta\rho}_{5i} \epsilon_\rho \left(p_\psi\right) \epsilon_\delta \left(p_{\rho}\right) , \\
\mathcal{M}_{6} &=& \epsilon^*_\mu \left(p_{D^*}\right)\epsilon^*_\nu
\left(p_{\bar D^*}\right)\sum_{i} \mathcal{M}^{\mu\nu\delta\rho}_{6i}\epsilon_\rho
\left(p_\psi\right) \epsilon_\delta \left(p_{\rho}\right)
\end{eqnarray}
where $\epsilon_\delta \left(p_{\rho}\right) $ is the polarization vector of the
$\rho$ meson. Again, the conjugate of $\mathcal{M}_{5}$ gives the amplitude for
the $\bar{D}^*D$ final state.
\begin{figure}[!htb]
\begin{center}
%\scalebox{0.55}[0.55]{\includegraphics{RHOJ.eps}}
\includegraphics[scale=0.60]{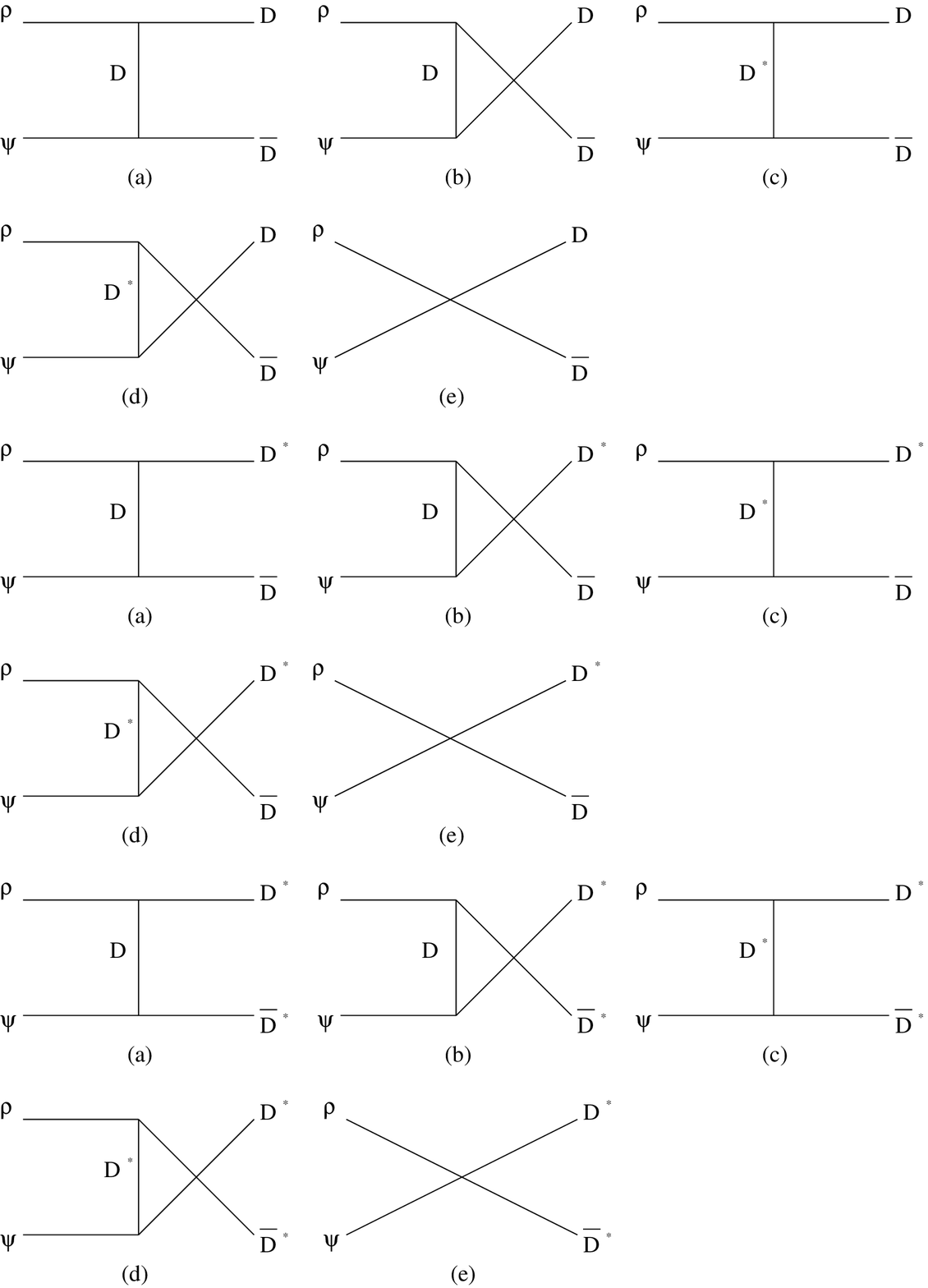}
\caption{Diagrams for $\rho +\psi$ dissociation}
\label{MRHOJ}
\end{center}
\end{figure}
It is important to note that the chiral symmetry constraint does not introduce
additional amplitudes involving the exchange of the $D_0^*$ and the $D_1$
 as in the case of the dissociation with pions. Consequently, the diagrams are 
the same as in Ref.~\cite{Oh01}, and we expect the results to agree.

\subsection{\label{softlimit_1}Soft--pion limit}
We now wish to demonstrate the soft--pion theorem for 
the dissociation of $J/\psi$ meson by a pion into a $D^*$--$\bar D$ final state. 
It is expected that this property of chiral
symmetry will soften the threshold behaviour. Explicitly, this will be due
 to a cancellation of the contact term  for the
normal--parity sub--processes.
The caveat here is of course that abnormal--parity interactions circumvent the
theorem and it will still be possible to have a contact behaviour near
the threshold due to these (Eq.~(\ref{M2b})).

With this in mind and in the chiral limit, i.e, for massless pions,
 we let the pion momentum go to zero. It is trivial to see
that the first sub--amplitude (Fig.~\ref{MPIJ}), due to the exchange of a $D$ meson
[Eq.~(\ref{M2a})] decouples when the vector mesons are on--shell since their
polarization vector then satisfies the orthogonality condition, i.e.,
$\epsilon(p) \cdot p = 0$. Similarly, the $u$--channel $D^*$ exchange amplitude goes to zero. 
We are thus left with three normal parity amplitudes including a contact term. In the soft--pion limit we
have
\begin{eqnarray}
\mathcal{M}_{2e} &\rightarrow& \frac{(2 \Delta^*)(2
g_{W\psi DD^*}\sigma_0)}{m_{D^*}^2
- m_{D_1}^2} g^{\mu \alpha} \nonumber \\
&\times&
\left\{g_{\alpha\beta} -\frac{p_{
D^*\alpha} p_{D^*\beta}}{m_{D_1}^{2}}\right\}g^{\beta
\rho}, 
\end{eqnarray}
and 
\begin{equation}
\mathcal{M}_{2f} \rightarrow \frac{(2 \Delta)(2 g_{W\psi DD^*}\sigma_0)}{m_D^2
- m_{D^*_0}^2} g^{\mu \rho}.
\end{equation}
Remembering that the the mass splittings between the $D$ mesons are due to
spontaneous chiral symmetry breaking, we can further write 
\begin{equation}
\mathcal{M}_{2e} \rightarrow -\frac{(2 \Delta)(2
g_{W\psi DD^*}\sigma_0)}{4\Delta^* \sigma_0}
g^{\mu \rho} = -g_{W\psi DD^*}g^{\mu \rho} 
\label{M2e_soft}
\end{equation}
and 
\begin{equation}
\mathcal{M}_{2f} \rightarrow -\frac{(2 \Delta)(2
g_{W\psi DD^*}\sigma_0)}{4\Delta \sigma_0} g^{\mu \rho}  = -g_{W\psi DD^*}g^{\mu \rho}
\label{M2f_soft}
\end{equation}
where for the amplitude $\mathcal{M}_{2e}$ the orthogonality condition has been
used to remove the term proportional to the product of four--vectors. Adding
these two contributions to the contact term of $\mathcal{M}_{2d}$ leads to the
desired result for the normal parity content. Since the contraction of the two $\epsilon$--tensors results
into a sum of products of the metric tensor, the leading behaviour near the threshold
for the process $\pi +\psi \rightarrow \bar D + D^*$ will be given by the amplitude $\mathcal{M}_{2b}$.
 
A remark is in order regarding the chiral limit.
Relaxing this assumption will make the amplitudes $\mathcal{M}_{2e}$ and
$\mathcal{M}_{2f}$ depend on the pion mass. It is trivial to see that these
can be mapped smoothly into the chiral amplitudes considered above by letting the pion
mass go to zero, thus satisfying the smoothness assumption of the decoupling
theorem.

\section{\label{cs_section}Cross--sections for dissociation processes}
\subsection{Introducing symmetry conserving form factors}
To complete the description of the phenomenological model, form factors must be
introduced to account for the sub--structure of mesons. A Lorentz--invariant three-point 
form factor is introduced, namely
\begin{equation}
\mathcal{F}^M_3(q^2) = \frac{\Lambda^2}{\Lambda^2 + \left|q^2 - m_M^2\right|}
\label{FF3}
\end{equation}
where $q^2$ is the virtuality, $m_M$ is the meson mass, and $\Lambda$ is the range parameter. 
%$M_0$ will be set to $2$ GeV to account for the fact that the exchanged mesons are $D$ mesons. 
The cutoff  parameter will be set to two different values, namely $1$ and $2$ GeV, as in
previous studies \cite{Lin00,Hag00,Oh01}. These can be justified by noting that the
typical hadronic scale is about $1$ GeV and the exchanged mesons, which are open charmed mesons here, have
masses of about $2$ GeV.  One could relax the universality condition by introducing a different cutoff parameter 
for each interaction, but the assumption of a common $\Lambda$ 
is a realistic first approximation because the exchanged mesons are all
$D$ mesons.

The astute reader will note that the coupling 
constants should strictly be defined at the point where the form factor is one,
i.e., $q^2 = m_M^2$. This is not the case for all the coupling values extracted 
in Ref.~\cite{Oh01} which are used here. Indeed, the three--point couplings involving a 
$\rho$ or a $J/\psi$ meson are evaluated with these particles at zero virtuality.
Nevertheless, 
it will still be assumed that the couplings extracted with the $\rho$ or the $J/\psi$ meson off--shell 
are the same as those on--shell.

A form factor for the four--point interactions is also introduced.
%In Ref.~\cite{Hag00} is it chosen in such a way that the amplitudes respect the 
%Ward identity. Rather, 
Here, a dipole form is chosen, namely,
\begin{equation}
\mathcal{F}_4(s,t) = \frac{\Lambda^2}{\Lambda^2 + \left|t - M_0^2\right|}
\frac{\Lambda^2}{\Lambda^2 + \left|u -M_0^2\right|}
\label{F3}
\end{equation}
where $s+t+u = m_1^2+m_2^2+m_3^2+m_4^2$ and $M_0$ is a mass scale. This latter
parameter is given by the average of the $D$ and $D^*$ masses, i.e., $M_0=1.94$
GeV. The four--point form factor is then equal to one when $t=u=M_0^2$. 
Strictly speaking the normalization, i.e., the coupling constant, is defined at this point.

The above discussion omits the constraint due to chiral symmetry. Indeed, some 
of the three--point form factors are determined 
by four--point form factors. This is the case for all three--point interactions
generated by underlying four--point interactions, i.e., which have a $W$-field
factor (see Appendix \ref{interactions} for details). Specifically, let us consider the $W\psi D D^*$ interaction from which 
the $\pi\psi D D^*$, $\psi D D_1$ and $\psi D^* D^*_0$ 
interactions are generated after spontaneous chiral symmetry breaking. The
three--point form factors can then be extracted from the four--point form factor 
by letting the pion momentum go to zero. Specifically, assume that the $D^*$ and $D$
mesons are off-- and on--shell, respectively, then setting the pion momentum to zero
yields the desired form factor for the $\psi D^* D^*_0$:
\begin{eqnarray}
\lim_{p_\pi\rightarrow 0} \mathcal{F}_4(s,t) &=& \frac{\Lambda^2}{\Lambda^2 + \left|m_{D^*}^2 - M_0^2\right|}
\frac{\Lambda^2}{\Lambda^2 + \left|u -M_0^2\right|}  \nonumber \\ &=& 
 \gamma_{D^*} \mathcal{F}^{0}_3(u)
\end{eqnarray}
where the index on the three--point form factor indicates that the
parameter $m_M$ is set to $M_0$, and  $\gamma_M = \mathcal{F}^0_3(m_M^2)$.
Taking the $D$ meson off--shell and keeping the $D^*$ on--shell gives the form factor 
for $\psi D D_1$ interaction. The same argument applies for the abnormal
 parity  $\psi D^* D_1$ and $\psi D^*_0 D$ interactions. 

There is also another subtlety when it comes to the interactions generating the
mass splittings of the $D$ mesons, i.e., those coming from $\mathcal{L}_{WDD}$ 
and $\mathcal{L}_{WD^*D^*}$. Indeed, the interaction form factors will now appear
in the mass shifts leading to self--consistent equations. For example, for the $D^*$--$D_1$ mass splitting,
we have
\begin{eqnarray}
m_{D_1}^2 -m_{D^*}^2 &=& 2\Delta^*\sigma_0 \lim_{p_\pi \rightarrow 0} 
\left\{\mathcal{F}^{D_1}_3(q^2) + \mathcal{F}^{D^*}_3(q^2)\right\} \nonumber \\
&=& 4\Delta^*\sigma_0 \frac{\Lambda^2}{\Lambda^2 + \left|m_{D_1}^2 -
 m_{D^*}^2\right|}\,.
\label{mass_split_Ds_2} 
\end{eqnarray}
From these, we see that the values of the interaction strengths, $\Delta^*$ and $\Delta$, are functions of both the cutoff parameter and the mass scale.

In light  of these modifications, we re--examined the soft--pion limit for 
the $\mathcal{M}_2$ amplitude. $\mathcal{M}_d$ is now given by
\begin{eqnarray}
\lim_{p_\pi \rightarrow 0}\mathcal{M}_{2d}
&=&2g_{W\psi DD^*}\gamma_{D^*}\gamma_{D}g^{\mu \rho}
\end{eqnarray}
while $\mathcal{M}_e$ and $\mathcal{M}_f$ reduce to 
\begin{eqnarray}
\lim_{p_\pi \rightarrow 0}\mathcal{M}_{\{2e,2f\}} &=& \lim_{p_\pi \rightarrow 0} 
\left[\gamma_D\mathcal{F}^{0}_3(t)\mathcal{F}_3^{D_1}(t)\right]\nonumber \\
&\times& \frac{(2 \Delta^*)(2
g_{W\psi DD^*}\sigma_0)}{m_{D^*}^2
- m_{D_1}^2} g^{\mu \rho} \nonumber \\
&= &-\gamma_D\gamma_{D^*}g_{W\psi DD^*}g^{\mu \rho}
\end{eqnarray}
where Eq.~(\ref{mass_split_Ds_2}) has been used to go from the first line to
the second.

\subsection{Results}
\begin{figure*}[!htb]
%\begin{center}
%\hspace{-2cm}
\includegraphics[scale=0.65]{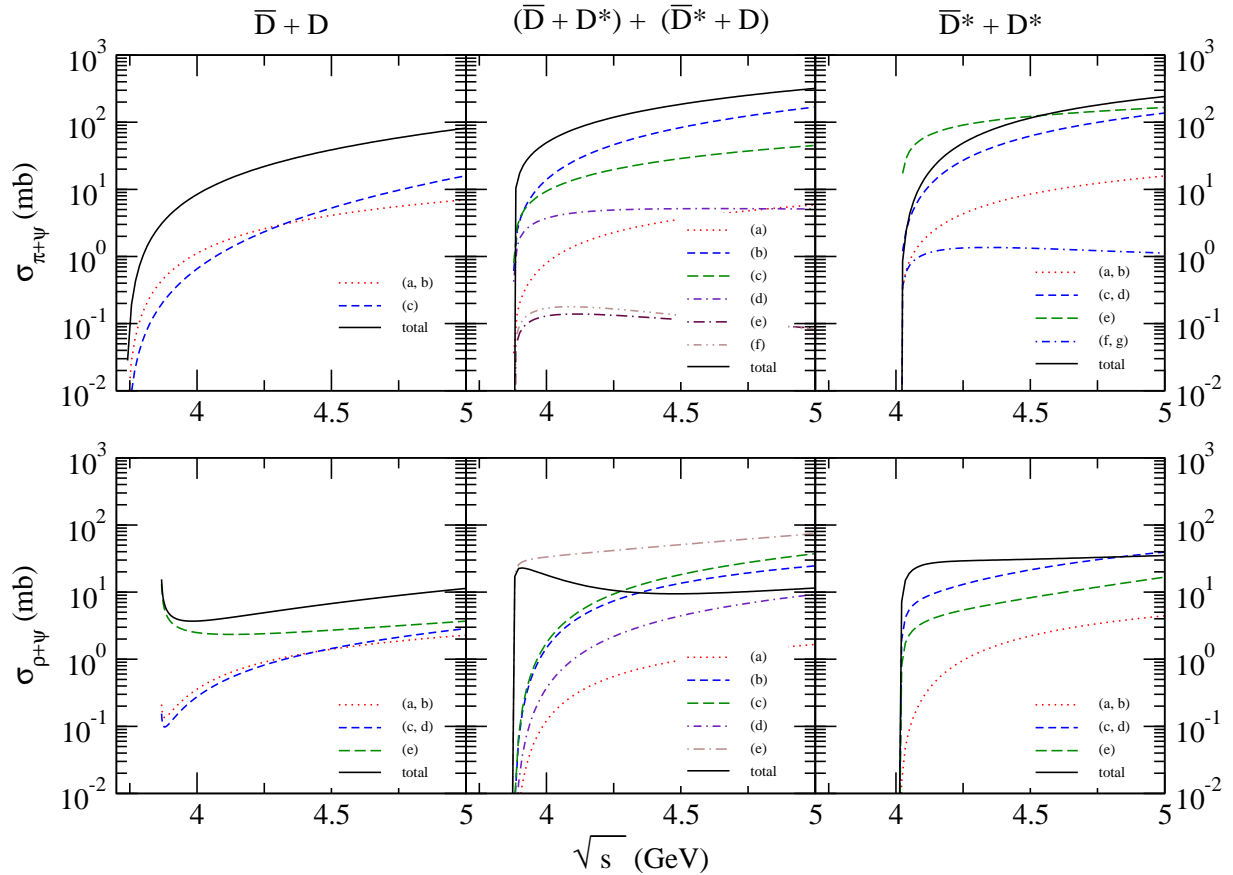}
\caption{\label{LCS_no_FF}(Color online) Dissociation cross--sections without form factors.}
%\end{center}
\end{figure*}

The cross--sections are first studied without form factors. The parameters used in the calculation
can be found in the Appendix \ref{parameters}. The six
cross--sections are presented in Fig.~\ref{LCS_no_FF} where the solid curves
are the cross--sections including all sub--amplitudes. Overall, near threshold both dissociation
by a pion and by a $\rho$ meson are of the 
same order of magnitude; the pion--dissociation starts to dominate over the
$\rho$--dissociation beyond $4$ GeV.

\begin{figure*}[!htb]
%\begin{center}
\includegraphics[scale=0.60]{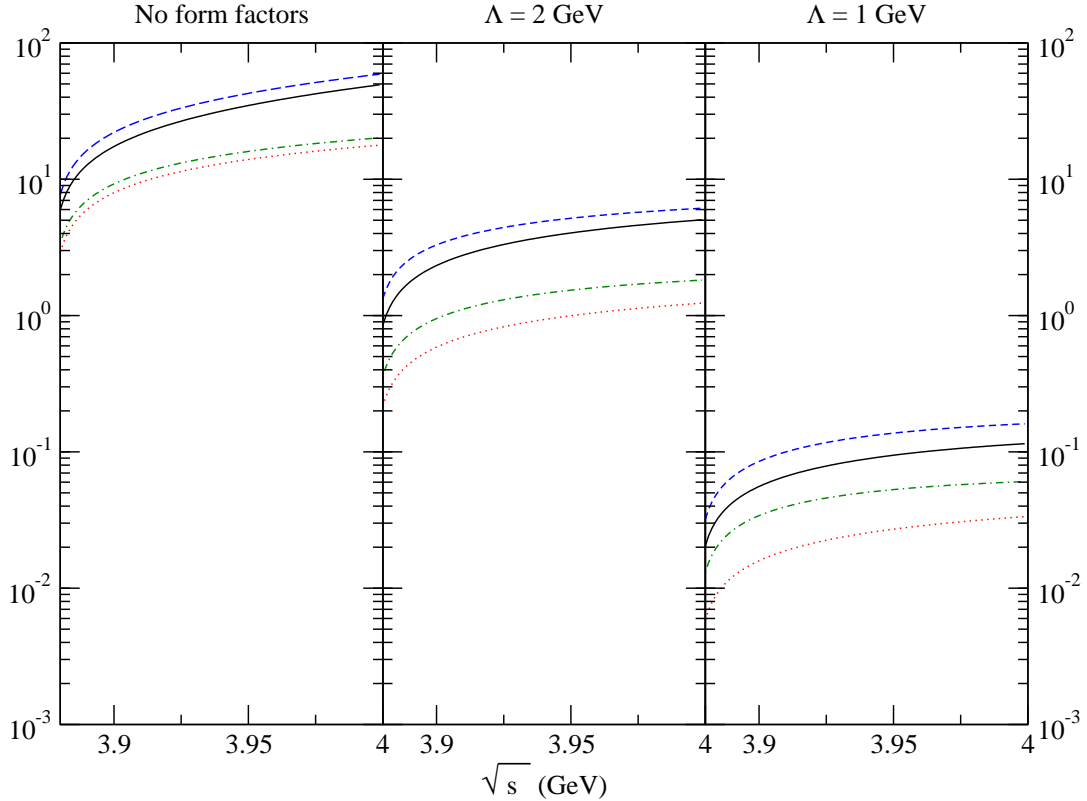}
\caption{\label{LCS_comparaison}(Color online) Effects of chiral symmetry and abnormal parity content on the $\pi + J/\psi \rightarrow (\bar D +D^*) + (\bar D^* + D)$ cross--section. The dotted, dashed, and
dot--dashed lines correspond to cross--sections without the abnormal parity
sub--amplitude,
without the two sub--amplitudes due to chiral symmetry, and without all three
sub--amplitudes.
The total inclusive cross--section with all contributions is given by the solid lines.}
%\end{center}
\end{figure*}

\begin{figure*}[!htb]
%\begin{center}
%\hspace{-1.0cm}
%\vspace{-0.5cm}
\includegraphics[scale=0.65]{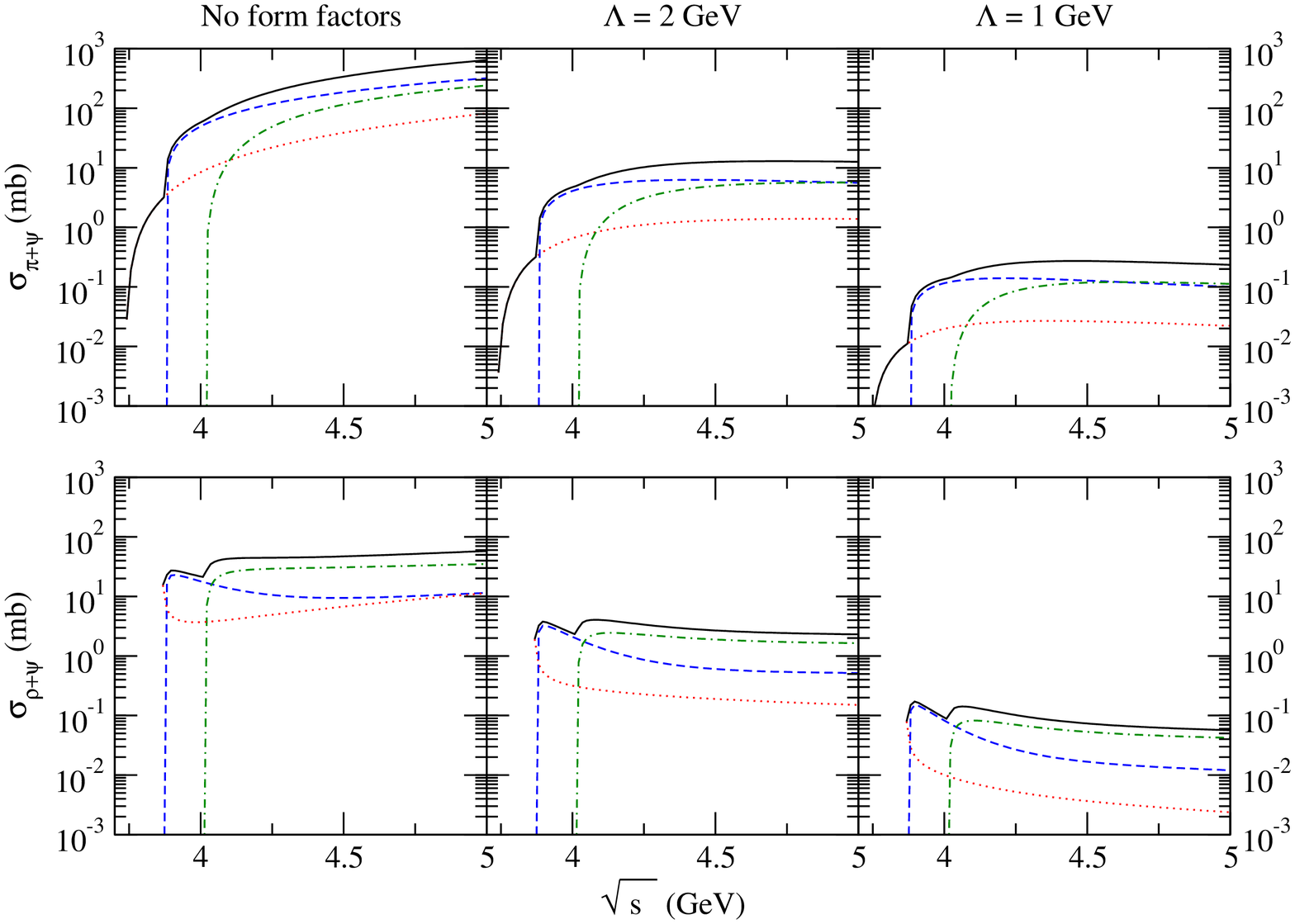}
\caption{\label{LCS_FF} (Color online) Comparison
of the dissociation cross--sections with
and without form factors. The dotted, dashed, and
dot--dashed lines correspond to dissociation into $\bar D + D$, $(\bar D +
D^*) + (\bar D^* + D)$, and $\bar D^* + D^*$. The total 
inclusive cross--sections are given by the solid lines; cusps are due to channels
opening.}
%
%\end{center}
\end{figure*}

The effect of introducing chiral symmetry can be assessed by considering 
the pion--absorption into the $(\bar D + D^*)+(\bar D^* +D)$ final state. The 
leading contribution to this process is due to the sub--amplitude $\mathcal{M}_{2b}$, 
which arises because of the abnormal parity content in the
Lagrangian. This is made clear in both Figs.~\ref{LCS_no_FF}
and \ref{LCS_comparaison}. Indeed, at a value of the centre--of--mass energy of $3.9$ GeV, 
excluding this sub--amplitude reduces the cross--section by $65\%$. In contrast, 
removing  $\mathcal{M}_{2e}$ and $\mathcal{M}_{2f}$,
i.e., the sub--amplitudes necessary to maintain chiral symmetry, 
only increases the cross--section by $27\%$. Although, omitting the chiral
constraint increases the cross--section, as expected from the soft--pion theorem, 
the effect is sub--leading compared to the inclusion of abnormal parity interactions.
Moreover, the presence of abnormal parity content makes the dissociation into 
$\bar D D$ and $\bar D^*D^*$ pairs possible, which further increases the total
pion--absorption cross--section.

Turning to $\rho$--dissociation, the initial expectation is that the results
of Ref.~\cite{Oh01} should be reproduced since no additional interactions are
introduced in applying the chiral symmetry constraint. Although, 
all three pion--absorption cross--sections are monotonically increasing 
with $\sqrt{s}$ and featureless, the three $\rho$--dissociation 
cross--sections differ qualitatively in shape, when  compared to results in  Ref.~\cite{Oh01}. 
In spite of the fact that the interactions and the squared sub--amplitudes are the same, the relative phases and, consequently,
the interference patterns are different leading to the observed dissimilarities. 

The above discussion is valid only when form factors are omitted. In
Fig.~\ref{LCS_FF}, cross--sections with and without form factors are compared. 
Two cases of the cutoff parameter are considered, namely $1$ and $2$ GeV.
As expected, for decreasing $\Lambda$ the suppression is increased. Overall, it is clear that the
magnitudes of the two dissociation channels are 
set by the functional forms of the form factors and the values of the parameters. 
With this caveat, the inclusive pion--dissociation cross--section is of the order of a few 
millibarns near threshold for a cutoff of $2$ GeV, and a fraction of millibarn
for $\Lambda =1$ GeV.

Finally, the relative effect of chiral symmetry as the cutoff 
is lowered is shown in Fig.~\ref{LCS_comparaison}. At
$\sqrt{s} = 3.9$ GeV, the cross--section for $\Lambda = 1$ GeV increases 
by $51\%$ when the sub--amplitudes due to chiral symmetry are neglected, 
while it decreases by $72\%$ when the abnormal parity content is omitted. 

\section{Conclusion and outlook}
We have presented an extension of the work done in Ref.~\cite{Bou05_1} 
where, besides introducing chiral symmetric interactions,
abnormal parity content and $\rho$ mesons are also included. The former is 
important since the soft--pion theorem is circumvented in this case, while the latter is a first step 
towards assessing the relative importance of the $J/\psi$--dissociation by other light resonances. To
account for the quark sub--structure of mesons, ad--hoc mesonic form factors were also added. Comparing the $\rho$--induced dissociation with the pion ones
did not shed any more light than what was found in Ref.~\cite{Oh01}. Any statements about the relative
strength between $\pi$-- and $\rho$--induced dissociation depend heavily on the choice of form factors and the techniques used to fix their absolute normalisations, and are thus model--dependent. 

We also conclude that there are some indications that the introduction of chiral symmetry does reduce
the cross--section of  $\pi + J/\psi \rightarrow (\bar D + D^*) + (D +
\bar D^*)$, but also that the implementation of abnormal parity content is probably as or even more important since it increases not
only the maximum reached by the $\pi + J/\psi \rightarrow (\bar D + D^*) + (D +
\bar D^*)$ cross--section, but also it allows new decay channels, such as $\pi + J/\psi \rightarrow \bar D^* +
D^*$, to open. 

In the future, the $J/\psi$--dissociation rates will be integrated in an evolving hot and dense medium. Introducing other light mesons, such as the $\omega$, as well as higher
charmonium resonances will also be considered to improve the phenomenological
description. Moreover, adding final states incorporating $D^*_0$ and
$D_1$  mesons and evaluating the cross--sections for the inverse
reactions will also figure with the additional developments. Then, contact with the phenomenology measured at the SPS and at RHIC will be made. 

\begin{acknowledgments}
This work was funded in part by the Natural Sciences and Engineering Research
Council of Canada, and by the Fonds Nature et Technologies of Quebec. A. Bourque
would like to thank K. L. Haglin for useful comments.
\end{acknowledgments}

\appendix
\section{\label{field_rep} Field representations and chiral symmetry}
In order to write down all the possible invariant interactions between the
mesons, it is essential to know their chiral transformation property.
Obviously, for the $J/\psi$ meson this is trivial as it is a singlet of the chiral
group. For the $\pi$, $\rho$, $D$, $D^*$, and their chiral partner, it is convenient to define 
chiral fields. 

The field representations of the $\pi$-- and $\sigma$--mesons are given by
\begin{eqnarray}
W = \sigma +i\pi\\
W^\dagger = \sigma -i\pi
\end{eqnarray}
Their transformation property 
under the $SU_L(2)\times SU_R(2)$ group can be assessed by coupling the 
chiral meson fields to quark bi--linears of corresponding parity giving
\begin{equation}
\bar q \left(\sigma + i\gamma_5 \pi \right) q
\end{equation}
where $\pi = \tau^a \pi^a$.  
Projecting the quark fields into  their left-- and right--handed representations
yields
\begin{equation}
\bar q \left(\sigma + i\gamma_5 \pi \right) q = \bar q_L W q_R + \bar
q_RW^\dagger q_L
\end{equation}
Under a chiral transformation of the light quark fields as defined by
\begin{equation}
q_{R,L} \rightarrow U_{R,L}q_{R,L} = e^{ -i\tau^i\epsilon^i_{R,L}} q_{R,L}
\end{equation}
where $\tau^i$ are the $SU(2)$ Pauli matrices satisfying the normalization
condition $Tr\left(\tau^i\tau^j\right) = 2\delta^{ij}$, the
chiral mesonic fields have to transform as 
\begin{eqnarray}
W \rightarrow U_L W U^\dagger_R\\
W^\dagger \rightarrow U_R W U^\dagger_L
\end{eqnarray}
for the interaction to be invariant. 

The spin--$1$ light mesons will not be
introduced as gauge bosons as in Ref.~\cite{Bou05_1}. Applying the same technique as for the $\sigma$ and
$\pi$ fields yields the interaction 
\begin{equation}
\bar q \left(\displaystyle{\not}\rho
 + \displaystyle{\not}a_1
 \gamma_5 \right)  q = 
\bar q_L A_L q_L + \bar q_R A_R q_R
\end{equation}
where now $ \rho_\mu = \rho^a_\mu \tau^a = \frac{1}{2}\left(A_R+A_L\right)$ 
and $a_{1\mu} = a^a_{1\mu}\tau^a = \frac{1}{2}\left(A_R-A_L\right)$. 
From these we infer that 
\begin{eqnarray}
A_\mu^{L} &\rightarrow& U_L A^L_\mu U^{\dagger}_L \\
A_\mu^{R} &\rightarrow& U_R A^R_\mu U^{\dagger}_R
\end{eqnarray}
which do not transform as gauge bosons.

Turning now to the open charmed mesons, we consider first the $D$ and $D^*_0$
isospin doublet fields and their conjugate which are written as 
\begin{eqnarray}
\bar D^T &=& \left(\bar D^0, D^-\right),\quad D = \left(D^0, D^+\right)
\nonumber \\
\bar D_0^{*T} &=& \left(\bar D_0^{*0}, D_0^{*-}\right),\quad D_0^* = \left(D_0^{*0},
D_0^{*+}\right)
\end{eqnarray}
where $T$ is the transposition operator. These can be re--arranged into 
\begin{eqnarray}
\bar D_{R,L} &=& \left(\bar D_0^* \mp i \bar D \right), \quad 
D_{R,L} = \left( D_0^* \pm i  D \right).
\end{eqnarray}
Since the open charmed mesons have only one light valence quark, they are expected
to transform under chiral symmetry according to 
\begin{eqnarray}
\bar D_{R,L} \rightarrow U_{R,L}\bar D_{R,L}  \nonumber \\ 
D_{R,L} \rightarrow  D_{R,L} U^\dagger_{R,L}
\label{D_RL_trans}
\end{eqnarray}
which can be made explicit by considering the coupling to the quark bi-linears:
\begin{eqnarray}
\bar Q \left( D_0^* + i D\gamma_5\right) q = \bar Q_L  D_R q_R +
\bar Q_R  D_L q_L
\nonumber \\
\bar q \left(  \bar D_0^* + i\bar D\gamma_5\right) Q = \bar q_L  \bar D_L Q_R +
\bar q_R \bar D_R Q_L
\end{eqnarray}
Similarly, the $D^*$ and $D_1$ fields can be cast into chiral forms yielding
\begin{eqnarray}
\bar D^*_{R,L} = \left(\bar D^* \pm \bar D_1 \right), \quad
D^*_{R,L} = \left( D^* \pm  D_1 \right) 
\end{eqnarray}
and the quark--meson interactions then read 
\begin{eqnarray}
\bar Q \left( \displaystyle{\not}D^* + \displaystyle{\not} D_1\gamma_5\right) q 
= \bar Q_L D^*_R q_R + \bar Q_R  D^*_L q_L \nonumber \\
\bar q \left( \displaystyle{\not}\bar D^* + \displaystyle{\not}\bar D_1\gamma_5\right) Q 
= \bar q_L \bar D^*_L Q_L + \bar q_R \bar D^*_R q_R
\end{eqnarray}
from which transformation properties similar to Eq.~(\ref{D_RL_trans}) are 
deduced.

\section{\label{interactions}Chiral invariant interactions}
Table \ref{C_P_trans} lists the chiral field properties under discrete transformations.
They are particularly useful to fix the relative signs of the interaction terms.
Moreover, the concepts of normal and abnormal parity interactions are also
introduced as a classification. Abnormal parity interactions have an
$\epsilon$--tensor factor. 
\begin{table*}[htbc] 
\begin{tabular}{|c|c|c|c|c|c|c|c|}
\hline
&$\psi^\mu (J/\psi)$ & $W(W^\dagger)$ & $D_{R,L}(\bar D_{R,L})$ &
$D^{*\mu}_{R,L}(\bar D^{*\mu}_{R,L})$ & $A^\mu_{R,L}$ & $\partial_\mu$&
$\epsilon^{\mu\nu\alpha\beta}$ \\
\hline 
$P$&$-\psi^\mu$&$W^\dagger(W)$&$D_{L,R}(\bar D_{L,R})$&$-D^{*\mu}_{L,R}(-\bar
D^{*\mu}_{L,R})$&$-A^\mu_{L,R}$&$-\partial_\mu$&$-\epsilon^{\mu\nu\alpha\beta}$\\
\hline
$C$&  $-\psi^\mu$ &$W^*(W^T)$&$\bar D^T_{R,L}(D^T_{R,L})$&$-\bar D^{*\mu
T}_{R,L}(-D^{*\mu T}_{R,L})$&$-A^{\mu T}_{R,L}$&$ +\partial_\mu$
&$-\epsilon^{\mu\nu\alpha\beta}$\\
\hline
\end{tabular}
\caption{\label{C_P_trans}Field transformation properties under parity and charge conjugation.}
\end{table*}

%\begin{widetext}
The three--point normal--parity interactions are then:
\begin{widetext}
\begin{eqnarray}
\mathcal{L}_{WDD} &=& -\Delta \left(D_L W \bar D_R+D_R W^\dagger \bar D_L\right)
\label{LWDD}
\\ 
\mathcal{L}_{WD^*D^*} &=& -\Delta^*\left(D^{*\mu}_L W \bar D^*_{R\mu}+D^{*\mu}_R W^\dagger \bar
D^*_{L\mu}\right) 
\label{LWDsDs}\\
\mathcal{L}_{WDD^*} &=& ig^{(0)}_{WDD^*}\left(\partial_\mu D_LW\bar D^{*\mu}_R + \partial_\mu D_R W^\dagger \bar
D^{*\mu}_L\right)  \nonumber \\
&+&ig^{(1)}_{WDD^*}\left(D_L\partial_\mu W\bar D^{*\mu}_R +  D_R \partial_\mu W^\dagger \bar
D^{*\mu}_L\right) \nonumber \\ &+& h.c.
\label{LWDDs} \\
\mathcal{L}_{\psi D^*D^*} &=& ig^{(0)}_{\psi D^*D^*}\partial_\mu\psi_\nu\left(D^{*\mu}_R\bar D^{*\nu}_R + D_L^{*\mu}\bar
D_L^{*\nu}\right) \nonumber \\
&+&ig^{(1)}_{\psi D^*D^*}\psi_\mu\left(\partial^\mu D^{*}_{R\nu}\bar D^{*\nu}_R + \partial^\mu
D_{L\nu}^*\bar D_L^{*\nu}\right) \nonumber \\
&+&ig^{(2)}_{\psi D^*D^*}\psi_\mu\left(\partial_\nu D^{*\mu}_R\bar D^{*\nu}_R + \partial_\nu
D_L^{*\mu}\bar D_L^{*\nu}\right) \nonumber \\ &+&h.c.,\\ 
\mathcal{L}_{A D^*D^*} &=& ig^{(0)}_{A D^*D^*}\left(D^{*\nu}_R\partial_\mu A_{R\nu}\bar D^{*\mu}_R +
D_L^{*\nu}\partial_\mu A_{L\nu}\bar D_L^{*\mu}\right) \nonumber \\
&+&ig^{(1)}_{A D^*D^*}\left( D^{*}_{R\nu} A_{R\mu} \partial^\mu\bar D^{*\nu}_R + 
D_{L\nu}^*A_{L\mu}\partial^\mu\bar D_L^{*\nu}\right) \nonumber \\
&+&ig^{(2)}_{A D^*D^*}\left(D^{*\mu}_R A_{R\nu}\partial_\mu\bar D^{*\nu}_R + 
D_L^{*\mu} A_{L\nu} \partial_\mu \bar D_L^{*\nu}\right) \nonumber \\ &+&h.c.,\\ 
\mathcal{L}_{\psi DD} &=& ig_{\psi DD}\psi_\mu\left(\partial^\mu D_R \bar D_R + \partial^\mu D_L \bar D_L \right
) +h.c.,\nonumber \\
\mathcal{L}_{A DD} &=& ig_{A DD}\left(D_R A^\mu_R\partial_\mu \bar D_R + D_LA_L^\mu\partial_\mu \bar
D_L\right) \nonumber \\ &+& h.c.
\end{eqnarray}
\end{widetext}
while the four--point interactions read
\begin{widetext}
\begin{eqnarray}
\mathcal{L}_{WWWW} &=&
-\frac{1}{16}\lambda^2\left(Tr\left[WW^\dagger\right]\right)^2, \\
\mathcal{L}_{W\psi D D^*} &=& g_{W\psi D D^*} \psi^\mu\left(D_L W \bar D^*_{R\mu} + D_R W^\dagger \bar
D^*_{L\mu} \right) \nonumber \\ &+& h.c., \\
\mathcal{L}_{A\psi D D} &=& g_{A\psi D D}\psi_\mu \left( D_R
A^\mu_R \bar D_{R} + D_L A^\mu_L \bar D_{L} \right)\\
\mathcal{L}_{A\psi D^* D^*} &=& g^{(0)}_{A\psi D^* D^*}\psi_\mu \left( D^{*\nu}_R
A^\mu_R \bar D^*_{R\nu} + D^{*\nu}_L A^\mu_L \bar D^*_{L\nu} \right) \nonumber
\\ &+& g^{(1)}_{A\psi D^* D^*}\psi_\mu \left( D^{*\nu}_R
A^\nu_R \bar D^{*R\mu} + D^{*\nu}_L A^\nu_L \bar D^{*L\mu} \right. \nonumber \\ 
&+&\left. h.c. \right) 
\end{eqnarray}
\end{widetext}
where $h.c.$ refers to the hermitian conjugate. All the coupling constants are
dimensionless with the exception of $\Delta$ and $\Delta^*$ which have 
dimension of mass. 

Abnormal--parity interactions cannot be written down directly at this point as there
remains an ambiguity in their definitions. Indeed, the 
interaction forms are not unique as there is a non--trivial relation 
called the Schouten's identity, relating different 
matrix elements \cite{Iva05,Kor03}. 
To build the interactions, the gauged Wess--Zumino Lagrangian
is used as a guide as in Ref.~\cite{Oh01}. The three--point interactions 
are then 
\begin{widetext}
\begin{eqnarray}
\mathcal{L}_{W D^* D^*} &=& ig_{W D^* D^*} \epsilon^{\mu\nu\alpha\beta}
\left(\partial_\mu  D^*_{L\nu}  W \partial_\alpha \bar D^*_{R\beta}  -
\partial_\mu  D^*_{R\nu}  W^\dagger \partial_\alpha \bar D^*_{L\beta} \right), \\
\mathcal{L}_{\psi D D^*} &=& ig_{\psi D D^*} \epsilon^{\mu\nu\alpha\beta}
\partial_\mu \psi_\nu \left(\partial_\alpha D^*_{L\beta} \bar D_{L} -
 \partial_\alpha D^*_{R\beta} \bar D_{R}\right) +h.c., 
 \label{LJDDs}\\
 \mathcal{L}_{A D D^*} &=& ig_{A D D^*} \epsilon^{\mu\nu\alpha\beta}
\left(\partial_\mu D^*_{L\nu} \partial_\alpha A_{L\beta} \bar D_{L} -
 \partial_\mu D^*_{R\nu} \partial_\alpha A_{R\beta} \bar D_{R}\right) +h.c.
\end{eqnarray}
\end{widetext}
 and the four--point interactions are given by
\begin{widetext}
\begin{eqnarray}
\mathcal{L}_{W\psi D D} &=& g_{W \psi D D} \epsilon^{\mu\nu\alpha\beta}
\psi_\mu\left(\partial_\nu D_L \partial_\alpha W \partial_\beta \bar D_R - 
\partial_\nu D_R \partial_\alpha W^\dagger \partial_\beta \bar D_L\right), \\
\mathcal{L}_{W\psi D^* D^*} &=& -g^{(0)}_{W \psi D^* D^*} \epsilon^{\mu\nu\alpha\beta}
\psi_\mu\left(D^*_{L\nu}\partial_\alpha W \bar D^*_{R\beta} - D^*_{R\nu}\partial_\alpha
W^\dagger \bar D^*_{L\beta}\right) \nonumber \\
&-&  g^{(1)}_{W \psi D^* D^*} \epsilon^{\mu\nu\alpha\beta}
\partial_\mu \psi_\nu\left(D^*_{L\alpha} W \bar D^*_{R\beta} - D^*_{R\alpha}
W^\dagger \bar D^*_{L\beta}\right) , \\
\mathcal{L}_{A\psi D D^*} &=& g^{(0)}_{A \psi D D^*} \epsilon^{\mu\nu\alpha\beta}
\psi_\mu \left( \partial_\nu D_R A_{R\alpha}\bar D^*_{R\beta} -
\partial_\nu D_L A_{L\alpha}\bar D^*_{L\beta} \right) \nonumber \\
&-& g^{(1)}_{A \psi D D^*} \epsilon^{\mu\nu\alpha\beta}
\psi_\mu \left( D_R A_{R\nu}\partial_\alpha \bar D^*_{R\beta} -
D_L A_{L\nu}\partial_\alpha \bar D^*_{L\beta} \right) + h.c.
\end{eqnarray}
\end{widetext}
where all couplings scale as $M^{-1}$ with the exception of $g_{W\psi DD}$ which
behaves as $M^{-3}$.

Once chiral symmetry is spontaneously broken the relevant normal--parity interactions become
\begin{widetext}
\begin{eqnarray}
\mathcal{L}_{\pi DD^*_0} &=& -2\Delta\left(D^*_0 \pi \bar D + D \pi \bar D^*_0
\right), \\
\mathcal{L}_{\pi D^*D_1} &=& -2\Delta^*i\left(D^*_\mu\pi\bar
D^\mu_1-D^\mu_1\pi\bar D^*_\mu\right), \\
\mathcal{L}_{\pi DD^*} &=& 2ig^{(0)}_{WDD^*} 
\left( \partial_\mu D \pi \bar D^{*\mu} - D^{*\mu}\pi \partial_\mu \bar D\right)
\nonumber \\
&+& 2ig^{(1)}_{WDD^*} 
\left(  D \partial_\mu \pi \bar D^{*\mu} - D^{*\mu} \partial_\mu \pi  \bar
D\right), \\ 
\mathcal{L}_{\psi D^*D^*} &=& 2ig^{(0)}_{\psi
D^*D^*}\partial_\mu\psi_\nu\left(D^{*\mu}\bar D^{*\nu} - D^{*\nu}\bar
D^{*\mu}\right) \nonumber \\
&+&2ig^{(1)}_{\psi D^*D^*}\psi_\mu\left(\partial^\mu D^{*}_{\nu}\bar D^{*\nu} - 
 D_{\nu}^*\partial^\mu\bar D^{*\nu}\right) \nonumber \\
&+&2ig^{(2)}_{\psi D^*D^*}\psi_\mu\left(\partial_\nu D^{*\mu}\bar D^{*\nu} - 
D^{*\nu} \partial_\nu\bar D^{*\mu}\right),\\ 
\mathcal{L}_{\rho D^*D^*} &=& 2ig^{(0)}_{A D^*D^*}\left(D^{*\nu}\partial_\mu
\rho_{\nu}\bar D^{*\mu} - D^{*\mu}\partial_\mu
\rho_{\nu}\bar D^{*\nu} \right)\nonumber \\
&+&2ig^{(1)}_{A D^*D^*}\left(  
D^{*}_{\nu} \rho_{\mu} \partial^\mu\bar D^{*\nu} - 
\partial^\mu D^{*}_{\nu} \rho_{\mu} \bar D^{*\nu} \right) \nonumber \\
&+&2ig^{(2)}_{A D^*D^*}\left(D^{*\mu} \rho_{\nu}\partial_\mu\bar D^{*\nu} -
\partial_\mu D^{*\nu} \rho_{\nu}\bar D^{*\mu} \right), \\
\mathcal{L}_{\psi DD} &=& 2ig_{\psi DD}\psi_\mu\left(\partial^\mu D \bar D - \partial^\mu D \bar D \right
), \\
\mathcal{L}_{\rho DD} &=& 2ig_{A DD}\left(D \rho^\mu \partial_\mu \bar D - 
\partial_\mu D \rho^\mu \bar D\right), \\
\mathcal{L}_{\psi D^*_0 D^*} &=& 2g_{W\psi D D^*} \sigma_0 \psi^\mu
\left(D_0^*\bar D^*_\mu + D^*_\mu \bar D_0^*\right), \\
\mathcal{L}_{\psi D D_1} &=& 2ig_{W\psi D D^*} \sigma_0 \psi_\mu
\left( D^\mu_1 \bar D - D\bar D^\mu_1 \right),
\end{eqnarray}
for three--point normal--parity interactions and
\begin{eqnarray}
\mathcal{L}_{\pi \psi DD^*} &=& 2g_{W\psi D D^*} \psi^\mu \left(D\pi\bar D^*_\mu +
D^*_\mu \pi \bar D\right),\\
\mathcal{L}_{\rho \psi DD} &=& 2g_{A\psi D D} \psi^\mu D\rho_\mu\bar D, \\ 
\mathcal{L}_{\rho \psi D^*D^*} &=& 2g^{(0)}_{A\psi D^* D^*} \psi^\mu
 D^{*\nu}\rho_\mu \bar D^*_\nu \nonumber \\
 &+& 2g^{(1)}_{A\psi D^* D^*} \psi^\mu\left(
 D^{*}_{\mu}\rho^\nu \bar D^*_\nu +  D^{*\nu}\rho_\nu \bar D^{*\nu}  \right).  
\end{eqnarray}
\end{widetext}
for the four--point normal--parity interactions. The last two three--point
interactions are induced from $\mathcal{L}_{W\psi D D^*}$. These play an
essential role in showing the decoupling of the pion from the dissociation
amplitude in the soft--momentum limit. As mentioned in Section \ref{densities}, the coupling constant $g^{(0)}_{WDD^*}$ is 
set to zero in order to remove the mixing between the various $D$ mesons. Furthermore, we drop the index on the remaining
coupling constant $g^{(1)}_{WDD^*} \rightarrow g_{WDD^*}$. For the sake of making more transparent 
the correspondence with Ref.~\cite{Oh01}, we further set $
g^{(0,2)}_{\{\psi,A\} D^* D^*} = -g^{(1)}_{\{\psi,A\} D^*
D^*} \rightarrow g_{\{\psi,A\} D^* D^*} $ and $g^{(i)}_{A \psi D^* D^*} 
\rightarrow g_{A \psi D^* D^*}$. 

Similarly, the abnormal parity content 
is 
\begin{widetext}
\begin{eqnarray}
\mathcal{L}_{\pi D^*D^*} &=& -2g_{WD^*D^*} \epsilon^{\mu\nu\alpha\beta} \partial_\mu D^*_\nu \pi
\partial_\alpha \bar D^*_\beta , \\
\mathcal{L}_{\psi D D^*} &=& -2g_{\psi D D^*} \epsilon^{\mu\nu\alpha\beta} 
\partial_\mu \psi_\nu \left(\partial_\alpha D^*_\beta \bar D + D \partial_\alpha
\bar D^*_\beta \right) , \\
\mathcal{L}_{\rho D D^*} &=& -2g_{A D D^*} \epsilon^{\mu\nu\alpha\beta} 
 \left(\partial_\alpha D^*_\beta \partial_\mu \rho_\nu \bar D + D \partial_\mu
 \rho_\nu
 \partial_\alpha \bar D^*_\beta \right), \\
 \mathcal{L}_{\psi D^* D_1} &=& 2g_{W\psi D^* D_1} \sigma_0\epsilon^{\mu\nu\alpha\beta} 
 \partial_\mu \psi_\nu \left( D_{1\alpha}  \bar D^*_\beta - D^*_\alpha \bar
 D_{1\beta} \right), 
\end{eqnarray}
\end{widetext}
where the last interaction is generated by $\mathcal{L}_{WA D^* D^*}$ and 
\begin{widetext}
\begin{eqnarray}
\mathcal{L}_{\pi \psi D D} &=& -2ig_{W\psi D D } \epsilon^{\mu\nu\alpha\beta} \psi_\mu \partial_\nu D
\partial_\alpha \pi \partial_\beta \bar D , \\
\mathcal{L}_{\pi \psi D^* D^*} &=& -2ig^{(0)}_{W \psi D^* D^* } \epsilon^{\mu\nu\alpha\beta} 
\psi_\mu D^*_\nu \partial_\alpha \pi \bar D^*_\beta \nonumber \\
&-& 2ig^{(1)}_{W\psi D^* D^* } \epsilon^{\mu\nu\alpha\beta} \partial_\mu
\psi_\nu D^*_\alpha \pi \bar D^*_\beta  ,\\
\mathcal{L}_{\rho \psi D^* D^*} &=& 2ig^{(0)}_{A \psi D D^* } \epsilon^{\mu\nu\alpha\beta} 
 \psi_\mu \left( \partial_\nu D \rho_\alpha  D^*_\beta + D^*_\nu \rho_\alpha
 \partial_\beta \bar D \right) \nonumber \\
 &-& 2ig^{(1)}_{A \psi D D^* } \epsilon^{\mu\nu\alpha\beta} 
 \psi_\mu \left( D \rho_\nu  \partial_\alpha D^*_\beta - \partial_\nu D^*_\alpha
  \rho_\beta \bar D \right),
\end{eqnarray}
\end{widetext}
for the three-- and four--point interactions. This completes the list of all
the required interactions.
\section{\label{amplitudes}Dissociation amplitudes}
\subsection{$\pi+J/\psi$}
We first investigate the dissociation process into two $D$ mesons illustrated in
 the first set of diagrams in Fig.~\ref{MPIJ}. The
sub--amplitudes are explicitly:
\begin{widetext}
\begin{eqnarray}
\mathcal{M}^{\rho}_{1a} &=& \frac{4g_{WDD^*}g_{\psi D D^*}}{t-m_D^{*2}}
p_\pi^\alpha \epsilon^{p_\psi p_{\bar D} \beta \rho} \nonumber \\
&\times&
\left\{g_{\alpha\beta}-\frac{\left(p_\pi-p_D\right)_\alpha\left(p_\pi-p_D\right)_\beta}{m_D^{*2}}\right\}				
, \\
\mathcal{M}^{\rho}_{1b} &=& -\frac{4g_{WDD^*}g_{\psi D D^*}}{u-m_D^{*2}}
p_\pi^\alpha \epsilon^{p_\psi p_{D} \beta \rho}\nonumber \\
&\times& \left\{g_{\alpha\beta} -\frac{\left(p_\pi-p_{\bar D}\right)_\alpha \left(p_\pi-p_{\bar
D}\right)_\beta}{m_D^{*2}}\right\}			
, \\
\mathcal{M}^{\rho}_{1c} &=& g_{W\psi D D } \epsilon^{p_\pi p_\psi p_{\bar D} \rho}
\end{eqnarray}
\end{widetext}
where $t=\left(p_\pi - p_D\right)^2$ and $u=\left(p_\pi - p_{\bar D}\right)^2$.
Note that there are no additional diagrams compared to Ref.~\cite{Oh01}.

Next we consider the absorption process which has been considered dominant in
the literature, namely $\pi +\psi \rightarrow \bar D + D^*$. As seen in
Fig.~\ref{MPIJ}, because of chiral
symmetry, the number of sub--processes is higher than in a theory where the chiral
partners are disregarded. Specifically, the list of sub--amplitudes for
this process is
\begin{widetext}
\begin{eqnarray}
\mathcal{M}^{\mu\rho}_{2a} &=& -\frac{4g_{WDD^*}g_{\psi D D}}{t-m_D^2}
p_\pi^\mu
\left(2p_{\bar D}^\rho - p_{\psi}^\rho\right),
\label{M2a} \\
\mathcal{M}^{\mu\rho}_{2b} &=& -\frac{4g_{WD^*D^*}g_{\psi D D^*}}{t-m_D^{*2}}
\epsilon^{p_\psi p_{D^*} \mu \alpha} \epsilon^{p_\psi p_{\bar D} \beta \rho}
%\nonumber \\ &\times& 
\left\{g_{\alpha\beta} -\frac{\left(p_\pi-p_{D^*}\right)_\alpha
\left(p_\pi-p_{D^*}\right)_\beta}{m_D^{*2}}\right\},\label{M2b} \\
\mathcal{M}^{\mu\rho}_{2c} &=&  -\frac{4g_{WD D^*}g_{\psi D^* D^*}}{{u-m_{D^*}^{2}}}
p_\pi^\alpha 
%\nonumber \\ &\times&
\left(2g^{\beta\rho}p_{\psi}^\mu - g^{\mu\rho}\left(p_\psi^\beta +
p_{D^*}^\beta\right)+2g^{\mu\beta} p_{D^*}^\rho \right) 
\nonumber \\ &\times&
\left\{g_{\alpha \beta} -\frac{\left(p_\pi-p_{\bar
D}\right)_\alpha\left(p_\pi-p_{\bar D}\right)_\beta}{m_D^{*2}}\right\},			
\label{M2c} \\
\mathcal{M}^{\mu\rho}_{2d} &=& g_{W\psi D D^* } g^{\mu\rho}, 
\label{M2d}\\
\mathcal{M}^{\mu\rho}_{2e} &=&  -\frac{4\Delta^* g_{W \psi D^*
D^*}\sigma_0}{t-m_{D_1}^{2}}
g^{\mu\alpha} g^{\beta\rho}
%\nonumber \\ &\times& 
\left\{g_{\alpha\beta} -\frac{\left(p_\pi-p_{
D^*}\right)_\alpha\left(p_\pi-p_{D^*}\right)_\beta}{m_{D_1}^{2}}\right\},
\label{M2e} \\ 
\mathcal{M}^{\mu\rho}_{2f} &=& \frac{4\Delta g_{W\psi D D^*}
\sigma_0}{u-m_{D^*_0}^{2}}
g^{\mu\rho}
\label{M2f}
\end{eqnarray}
\end{widetext}
where $t=\left(p_\pi - p_D^*\right)^2$ and $u=\left(p_\pi - p_{\bar
D}\right)^2$. We note that the the last two amplitudes arise because of the
exchange of the $D_1$ and $D_0^*$ mesons. 

Finally, the last pion--absorption process is that which leads
to the heaviest final state considered in this study, i.e., $D^*$--$\bar D^*$. The
sub--amplitudes related to the diagrams in Fig.~\ref{MPIJ} are  
\begin{widetext}
\begin{eqnarray}
\mathcal{M}^{\mu\nu\rho}_{3a} &=& \frac{4g_{WDD^*}g_{\psi D^* D}}{t-m_D^2}
p_\pi^\nu				
\epsilon^{p_\psi p_{\bar D^*} \mu \rho}, \\
\mathcal{M}^{\mu\rho}_{3b} &=& -\frac{4g_{WDD^*}g_{\psi D^* D}}{u-m_D^2}
p_\pi^\mu
\epsilon^{p_\psi p_{D^*} \nu \rho}, \\
\mathcal{M}^{\mu\nu\rho}_{3c} &=&  \frac{4g_{WD^* D^*}g_{\psi D^* D^*}}{{t-m_{D^*}^{2}}}
\epsilon^{p_\psi p_{D^*} \alpha \nu} 
%\nonumber \\ &\times&
\left(2g^{\beta\rho}p_{\psi}^\mu - g^{\mu\rho}\left(p_\psi^\beta +
p_{\bar D^*}^\beta\right)+2g^{\mu\beta} p_{\bar D^*}^\rho \right) \nonumber \\
&\times&
\left\{g_{\alpha\beta} -\frac{\left(p_\pi-p_{
D^*}\right)_\alpha\left(p_\pi-p_{ D^*}\right)_\beta}{m_D^{*2}}\right\}, \\
%\end{eqnarray}
%\begin{eqnarray}
\mathcal{M}^{\mu\nu\rho}_{3d} &=&  \frac{4g_{WD^* D^*}g_{\psi D^* D^*}}{{u-m_{D^*}^{2}}}
\epsilon^{p_\psi p_{\bar D^*} \alpha \mu} 
%\nonumber \\ &\times&
\left(2g^{\beta\rho}p_{\psi}^\nu - g^{\nu\rho}\left(p_\psi^\beta +
p_{D^*}^\beta\right)+2g^{\nu\beta} p_{ D^*}^\rho \right)
\nonumber \\ &\times&
\left\{g_{\alpha\beta} -\frac{\left(p_\pi-p_{\bar
D^*}\right)_\alpha\left(p_\pi-p_{\bar D^*}\right)_\beta}{m_D^{*2}}\right\}, \\
\mathcal{M}^{\mu\nu\rho}_{3e} &=& 2 g^{(0)}_{W\psi D^* D^* } \epsilon^{p_\pi\mu \nu \rho} 
+2g^{(1)}_{W\psi
D^* D^* }\epsilon^{p_{\psi}\mu \nu \rho}, \\
\mathcal{M}^{\mu\rho}_{3f} &=&  -\frac{4\Delta^* g_{W \psi D^*
D^*}\sigma_0}{t-m_{D_1}^{2}}
g^{\alpha\nu}\epsilon^{p_\psi\mu \beta \rho} 
%\nonumber \\ &\times&
\left\{g_{\alpha\beta} -\frac{\left(p_\pi-p_{
D^*}\right)_\alpha\left(p_\pi-p_{D^*}\right)_\beta}{m_{D_1}^{2}}\right\}, \\ 
\mathcal{M}^{\mu\rho}_{3g} &=& -\frac{4\Delta^* g_{W \psi D^*
D^*}\sigma_0}{u-m_{D_1}^{2}}
g^{\alpha\mu}\epsilon^{p_\psi\beta \nu \rho} 
%\nonumber \\&\times& 
\left\{g_{\alpha\beta} -\frac{\left(p_\pi-p_{
\bar D^*}\right)_\alpha\left(p_\pi-p_{\bar D^*}\right)_\beta}{m_{D_1}^{2}}\right\}
\end{eqnarray}
\end{widetext}
where $t=\left(p_\pi - p_{D^*}\right)^2$ and $u=\left(p_\pi - p_{\bar
D^*}\right)^2$. 

\subsection{$\rho+J/\psi$}
The amplitudes for the dissociation into the lowest mass state given in
Fig.~\ref{MRHOJ} are 
\begin{widetext}
\begin{eqnarray}
\mathcal{M}^{\delta\rho}_{4a} &=& -\frac{4g_{ADD}g_{\psi D D}}{t-m_D^2}
\left(2p_D^\delta - p_{\rho}^\delta\right)
%\nonumber \\ &\times& 
\left(2p_{\bar D}^\rho - p_{\psi}^\rho\right) \\
\mathcal{M}^{\delta\rho}_{4b} &=& -\frac{4g_{ADD}g_{\psi D D}}{u-m_D^2}
\left(2p_{\bar D}^\delta - p_{\rho}^\delta\right) 
%\nonumber \\ &\times& 				
\left(2p_{D}^\rho - p_{\psi}^\rho\right), \\
\mathcal{M}^{\delta\rho}_{4c} &=& -\frac{4g_{AD D^*}g_{\psi D D^*}}{t-m_D^{*2}}
\epsilon^{p_\rho p_{D} \alpha \delta}
\epsilon^{p_\psi p_{\bar D} \beta \rho} 
%\nonumber \\ &\times& 	
\left\{g_{\alpha\beta} -\frac{\left(p_\rho-p_{D^*}\right)_\alpha
\left(p_\rho-p_{D^*}\right)_\beta}{m_D^{*2}}\right\}, \\
\mathcal{M}^{\delta \rho}_{4d} &=& -\frac{4g_{AD D^*}g_{\psi D D^*}}{u-m_D^{*2}}
\epsilon^{p_\rho p_{\bar D} \alpha \delta}\epsilon^{p_\psi p_{D} \beta \rho}
%\nonumber \\ &\times& 			
\left\{g_{\alpha\beta} -\frac{\left(p_\rho-p_{\bar D}\right)_\alpha
\left(p_\rho-p_{\bar D}\right)_\beta}{m_D^{*2}}\right\}, \\ 
\mathcal{M}^{\delta\rho}_{4e} &=& -2g_{A\psi D D }g^{\delta\rho}.
\end{eqnarray}
\end{widetext}
where $t=\left(p_\rho - p_D\right)^2$ and $u=\left(p_\rho - p_{\bar
D}\right)^2$. Of all the six processes studied, this is the only one that is exothermic, i.e., the
initial state is more massive than the final one. This kinematical constraint
will give rise to a divergent cross--section behaviour at low $\sqrt{s}$.

The amplitudes of the second process (depicted in Fig.~\ref{MRHOJ}) are 
\begin{widetext}
\begin{eqnarray}
\mathcal{M}^{\mu\delta\rho}_{5a} &=& \frac{4g_{ADD^*}g_{\psi D D}}{t-m_D^2}
\epsilon^{p_\rho p_{ D^*} \mu \delta}
\left(2p_{\bar D}^\rho - p_{\psi}^\rho\right), \\				
\mathcal{M}^{\mu\delta\rho}_{5b} &=& \frac{4g_{ADD^*}g_{\psi D D}}{u-m_D^2}
\left(2p_{\bar D}^\delta - p_{\rho}^\delta\right)				
\epsilon^{p_\psi p_{ D^*} \mu \rho}, \\
\mathcal{M}^{\mu\delta\rho}_{5c} &=&  \frac{4g_{AD^* D^*}g_{\psi D D^*}}{{t-m_{D^*}^{2}}}
\left(2g^{\alpha \delta} p_\rho^\mu -g^{\mu\delta}\left(p_\rho^\alpha +p_{D^*}^\alpha\right)
+2g^{\alpha \mu}p_{D^*}^\delta \right) \nonumber \\
&\times& \left\{g_{\alpha\beta} -\frac{\left(p_\rho-p_{
D^*}\right)_\alpha\left(p_\rho-p_{ D^*}\right)_\beta}{m_D^{*2}}\right\}			
e^{p_\psi p_{\bar D} \beta \rho}
, \\
\mathcal{M}^{\mu\delta\rho}_{5d} &=&  \frac{4g_{AD D^*}g_{\psi D^* D^*}}{{u-m_{D^*}^{2}}}
\epsilon^{p_\rho p_{\bar D } \alpha \delta}
\left\{g_{\alpha\beta} -\frac{\left(p_\rho-p_{\bar
D}\right)_\alpha\left(p_\rho-p_{\bar D}\right)_\beta}{m_D^{*2}}\right\}			
\nonumber \\ 
&\times&\left(2g^{\beta\rho}p_{\psi}^\mu - g^{\mu\rho}\left(p_\psi^\beta +
p_{D^*}^\beta\right)+2g^{\mu\beta} p_{ D^*}^\rho \right), \\
\mathcal{M}^{\mu\delta\rho}_{5e} &=& 2 g^{(0)}_{A\psi D D^* } \epsilon^{p_{\bar
D}\mu \delta \rho} +2g^{(1)}_{A\psi D D^* }\epsilon^{p_{D^*}\mu \delta \rho}
\end{eqnarray}
\end{widetext}
where $t=\left(p_\rho - p_{D^*}\right)^2$ and $u=\left(p_\rho - p_{\bar D}\right)^2$. 

And finally, the set of amplitudes for the final dissociation processes, given in
Fig.~\ref{MRHOJ}, have the corresponding expressions:
\begin{widetext}
\begin{eqnarray}
\mathcal{M}^{\mu\nu\delta\rho}_{6a} &=& -\frac{4g_{ADD^*}g_{\psi D D^*}}{t-m_D^2}
\epsilon^{p_\rho p_{ D^*} \mu \delta}
\epsilon^{p_\psi p_{\bar D^*} \nu \rho}, \\
\mathcal{M}^{\mu\nu\delta\rho}_{6b} &=& -\frac{4g_{ADD^*}g_{\psi D D^*}}{u-m_D^2}
\epsilon^{p_\rho p_{ \bar D^*} \nu \delta}
\epsilon^{p_\psi p_{ D^*} \mu \rho}, \\
\mathcal{M}^{\mu\nu\delta\rho}_{6c} &=&  -\frac{4g_{AD^* D^*}g_{\psi D^* D^*}}{{t-m_{D^*}^{2}}}
\left\{g_{\alpha\beta} -\frac{\left(p_\rho-p_{
D^*}\right)_\alpha\left(p_\rho-p_{ D^*}\right)_\beta}{m_D^{*2}}\right\} \nonumber
\\
&\times&
\left(2g^{\alpha \delta} p_\rho^\mu -g^{\mu\delta}\left(p_\rho^\alpha +p_{D^*}^\alpha\right)
+2g^{\alpha \mu}p_{D^*}^\rho\right) \nonumber \\
&\times&
\left(2g^{\beta \rho} p_\rho^\nu -g^{\nu\rho}\left(p_\rho^\beta
+p_{\bar D^*}^\beta\right) +2g^{\beta \nu}p_{\bar D^*}^\rho\right), \\
\mathcal{M}^{\mu\nu\delta\rho}_{6d} &=&   -\frac{4g_{AD^* D^*}g_{\psi D^* D^*}}{t-m_{D^*}^{2}}
\left\{g_{\alpha\beta} -\frac{\left(p_\rho-p_{
D^*}\right)_\alpha\left(p_\rho-p_{ D^*}\right)_\beta}{m_D^{*2}}\right\} \nonumber
\\ 
&\times&
\left(2g^{\alpha \delta} p_\rho^\nu -g^{\nu\delta}\left(p_\rho^\alpha +p_{\bar D^*}^\alpha\right)
+2g^{\alpha \nu}p_{\bar D^*}^\rho\right) \nonumber \\
&\times&
\left(2g^{\beta \rho} p_\rho^\nu -g^{\nu\rho}\left(p_\rho^\beta
+p_{D^*}^\beta\right) +2g^{\beta \nu}p_{D^*}^\rho\right),\\
\mathcal{M}^{\mu\nu\delta\rho}_{6e} &=& g^{(0)}_{A\psi D^*
D^*} \left(2g^{\mu\nu}g^{\delta \rho} -g^{\mu\delta}g^{\nu \rho} - g^{\mu\rho}g^{\nu \delta} \right)
\end{eqnarray}
\end{widetext}
where $t=\left(p_\rho - p_{D^*}\right)^2$ and $u=\left(p_\rho - p_{\bar D^*}\right)^2$. 

\section{\label{parameters}Parameter fixing}
The coupling constants used here are fixed to those of Ref.~\cite{Oh01}.
There, besides fitting the available experimental data, 
they invoked the vector meson dominance hypothesis, the heavy quark
spin--flavor symmetry, and the underlying $SU(4)$ symmetry on which the
Lagrangian is built. Each of these assumptions is problematic. Unfortunately,
because experimental data are lacking to fix, for example,
the four--point couplings, the only other way would be to use other model calculations
with varying degree of sophistication. Table \ref{coupling_constants} lists the coupling
constant values used.
\begin{table}[htb] 
\begin{center}
\begin{tabular}{|c|c|c|c|}
\hline
\multicolumn{2}{|c|}{Three--point couplings} & \multicolumn{2}{|c|}{Four--point couplings} \\
\hline 
$g_{WDD^*}$&$4.40$&$g_{\psi D D^*}$&$16.96$\\
\hline 
$g_{\psi D^*D^*}$&$3.86$&$g_{A \psi D D}$&$19.43$\\
\hline 
$g_{\psi DD}$&$3.86$&$g_{A \psi D^* D^*}$&$9.72$\\
\hline 
$g_{A D^*D^*}$&$1.26$&$g_{W\psi DD}$&$8.00\,GeV^{-3}$\\
\hline 
$g_{A DD}$&$1.26$&$g^{(i)}_{W\psi D^*D^*}$&$19.10\,GeV^{-1}$\\
\hline 
$g_{W D^*D^*}$&$4.54\, GeV^{-1}$&$g^{(i)}_{A\psi D D^*}$&$10.89\,GeV^{-1}$\\
\hline 
$g_{\psi D D^*}$&$4.32\, GeV^{-1}$&&\\
\hline 
$g_{A D D^*}$&$1.41 \, GeV^{-1}$&&\\
\hline
\end{tabular}
\end{center}
\caption{\label{coupling_constants}Coupling constants of the phenomenological Lagrangian.}
\end{table}

Setting the coupling constants to those of Ref.~\cite{Oh01} is not sufficient to
determine all the parameters. Five parameters : $M$, $M^*$,
$\Delta$, $\Delta^*$, and $\sigma_0$  have to be determined. The last one is the decay constant, $f_\pi = 93$ MeV. The four remaining parameters 
have to reproduce the masses \footnote{For the $D$ and $D^*$ the isopin averaged
masses are used.} of the $D$, $D^*$, $D^*_0$, and $D_1$ mesons, 
namely $m_{D} = 1.87$ GeV, $m_{D^*} = 2.01$ GeV, 
$m_{D^*_0} = 2.40$ GeV,  and $m_{D_1} = 2.43$ GeV, respectively \cite{Yao06}. This leads to values 
of $M=2.15$ GeV and $M^* = 2.23$ GeV. Table \ref{couplingsII} lists the values of  $\Delta$ and
$\Delta^* $, and $\gamma_D$ and $\gamma_{D^*}$ used.
 Finally, the pion, $\rho$, and $J/\psi$ masses are taken to be $0.138$ GeV, $0.770$ GeV and $3.10$
GeV respectively.

\begin{table}[htb] 
\begin{center}
\begin{tabular}{|c|c|c|c|c|}
\hline
& $\Delta$ & $\Delta^*$ & $\gamma_{D}$ & $\gamma_{D^*}$\\
& (GeV) & (GeV) & & \\
\hline 
No Form Factor & $6.10$ & $5.01$ & $1$ & $1$\\
\hline 
$\Lambda = 1$ GeV & $19.85$ & $14.36$ & $0.79$ & $0.78$ \\
\hline 
$\Lambda = 2$ GeV & $9.53$ & $7.35$ & $0.94$ & $0.94$ \\
\hline 
\end{tabular}
\end{center}
\caption{\label{couplingsII}Cutoff--dependent coupling constants.}
\end{table}

%\begin{acknowledgments}
%\end{acknowledgments}

\bibliography{Lagrangian_final.bib}

\end{document}